\documentclass[prb,twocolumn,showpacs,preprintnumbers,amsmath,amssymb,floatfix,reprint]{revtex4-1}

\usepackage{color}
\usepackage{graphicx}
\usepackage{longtable}
\usepackage{dcolumn}
\usepackage{bm}
\usepackage{fancybox}

\newcommand{\GW}{$GW$}

\newcommand{\bfzero}{{\bf 0}}
\newcommand{\bfq}{{\bf q}}
\newcommand{\bfk}{{\bf k}}
\newcommand{\bfr}{{\bf r}}

\newcommand{\bfG}{{\bf G}}
\newcommand{\bfR}{{\bf R}}

\newcommand{\ispone}{}
\newcommand{\isptwo}{}

\newcommand{\req}[1]{\mbox{Eq.~\!(\ref{#1})}}
\newcommand{\refsec}[1]{\mbox{Sec.~\!\ref{#1}}}

\def\G0{G^0}

\def\QSGW{QS{GW}}

\def\phidot{\dot{\phi}}

\def\ei{\varepsilon_i}

\def\ej{\varepsilon_j}

\def\Psikn{\Psi_{{\bf k}n}}
\def\Psiqn{{\Psi_{{\bf q}n}}}
\def\Psiqm{{\Psi_{{\bf q}m}}}

\def\veff{V^{\rm eff}}
\def\vxc{V^{\rm xc}}
\def\Dvxc{{\it \Delta} V^{\rm xc}}

\def\vext{V^{\rm ext}}
\def\hVext{\hat{V}^{\rm ext}}
\def\hVeff{\hat{V}^{\rm eff}}

\def\vh{V^{\rm H}}

\def\hVee{\hat{V}^{\rm ee}}
\def\hVext{\hat{V}^{\rm ext}}
\def\hHk{\hat{H}^{\rm k}}

\def\Psiqkm{{\Psi_{{\bf q}-{\bf k}m}}}
\def\Psikn{{\Psi_{{\bf k}n}}}
\def\Psikm{{\Psi_{{\bf k}m}}}

\def\Psiqnp{{\Psi_{{\bf q}n'}}}
\def\Psiqn{{\Psi_{{\bf q}n}}}
\def\brl{{\bf R}L}

\def\connect#1{\leavevmode{\setbox1=\hbox{#1}\copy1%
\raise .2\ht1 \vbox{\moveleft \wd1\vbox{\hrule width \wd1 height .5pt depth 0pt}}%
}}

\def\ftn[#1]{\rlap{\footnotemark[#1]}}

\def\EMAX{  E^{\rm APW}_{\rm MAX} }

\def\EMAXS{E_{\rm MAX}^\Sigma}

\def\EMAX{  E^{\rm APW}_{\rm MAX} }

\def\ERPA{E^{\rm RPA}}
\def\bfp{\bf p}
\def\EMAX{  E^{\rm APW}_{\rm MAX} }
\def\H0{H^0}

\def\hH{\hat{H}}

\begin{document}
\title{Quasiparticle self-consistent GW method based on the augmented plane-wave and muffin-tin orbital method}
\author{Takao Kotani}
\affiliation{Department of applied mathematics and physics, Tottori university, Tottori 680-8552, Japan}
\date{\today}

\begin{abstract}
We have developed the quasiparticle self-consistent $GW$ (\QSGW) method based
on a recently developed mixed basis all-electron full-potential
method (the PMT method), which uses the augmented plane waves (APWs) and
the highly localized muffin-tin orbitals (MTOs) simultaneously. We call this PMT-QSGW.
Because of the two kinds of augmented bases, 
we have efficient description of
one-particle eigenfunctions in materials with small number of basis functions.
In QSGW, we have to treat a static non-local exchange-correlation potential,
which is generated from the self-energy. 
We expand the potential in the
highly localized MTOs.
This allows us to make stable interpolation of 
the self-energy in the whole Brillouin zone.
In addition, we have improved the offset-$\Gamma$ method for
the Brillouin zone integration, so that we take into account 
the anisotropy of the screened Coulomb interaction in the calculation of
the self-energy.
For GaAs and cubic SiO$_2$, we checked convergence 
of calculated band gaps on cutoff parameters.
PMT-QSGW is implemented in a first-principles electronic structure 
package \texttt{ecalj}, which is freely available from github.
\end{abstract}
\maketitle

\section{introduction }
The quasiparticle self-consistent $GW$ method (\QSGW)
is a self-consistent perturbation method within the $GW$ approximation.
\QSGW\  find out an optimum 
static one-body Hamiltonian $\hH^0$ describing the independent-particle picture
(or the quasiparticle (QP) picture).
In other words, \QSGW\ divides the full many-body Hamiltonian 
$\hH$ into $\hH=\hH^0+(\hH-\hH^0)$.
Then $(\hH-\hH^0)$ is chosen so that it 
virtually does not affect to the determination of QPs.
That is, we extract $\hH^0$ as a kernel of $\hH$.
Note that $(\hH-\hH^0)$ should contain not only the bare 
Coulomb interaction but also quadratic term,
which is missing in usual model Hamiltonians.
Since we evaluate $(\hH-\hH^0)$ in the $GW$ approximation in QSGW,
we determines $\hH^0$ (or the QPs, equivalently)
with taking into account the charge fluctuation 
in the random phase approximation (RPA) self-consistently.
\QSGW\ is conceptually completely different from the fully self-consistent $GW$ method 
\cite{holm_fully_1998,ku_band-gap_2002,stan_fully_2006,
PhysRevB.81.085103,PhysRevB.88.075105},
which tries to calculate the full one-body Green's function self-consistently.

\QSGW\ was first introduced by Faleev, 
van Schilfgaarde, and Kotani \cite{faleev_all-electron_2004}.
It was implemented based on the all-electron full-potential linearized 
muffin-tin orbital (FP-LMTO) package \cite{lmfchap} 
organized by van Schilfgaarde, in combination with 
the $GW$ package developed by Kotani initially for Ref.\onlinecite{kotani_all-electron_2002},
where he starts from a detailed analysis of a $GW$ package developed
by Aryasetiawan 
\cite{aryasetiawan_electronic_1994,aryasetiawan_product-basis_1994,aryasetiawan__1998} based 
on the LMTO in the atomic sphere approximation.
We refer to the implementation as FP-LMTO-QSGW in the followings.
\QSGW\ is now widely accepted as a possible candidate 
to go beyond limitations of current first-principles methods
\cite{vans06,kotani07a}
(we recently found that FP-LMTO-QSGW is taken to be 
for massively parallelized \cite{qsgwmass}).
\QSGW\ is also implemented in other first-principles electronic structure 
packages in different manners
\cite{hamann09,shishkin_accurate_2007,shaltaf_band_2008,bruneval_gw_2009,bruneval_ionization_2012,PhysRevB.84.205415}.
For example, Bruneval have calculated ionization energies of atoms in QSGW
\cite{bruneval_gw_2009,bruneval_ionization_2012}.

In \QSGW, we have to treat a  static non-local exchange-correlation potential $\vxc(\bfr,\bfr')$ (spin index is omitted for simplicity here). 
It is given by removing the energy-dependence from the self-energy $\Sigma(\bfr,\bfr',\omega)$ in a manner (See \req{eq:vxceq}). 
We determine eigenvalues and eigenfunctions with $\vxc(\bfr,\bfr')$, with which we evaluate not only the diagonal , but also the off-diagonal elements.
The importance of the off-diagonal elements is seen especially in the dispersion crossing. 
We see the (conventional) one-shot \GW\ only with the diagonal self-energy can not give a band gap for Ge as shown Fig.6 in Ref.\onlinecite{schilfgaarde_adequacy_2006}.
This is because the connectivity of the dispersion in GGA/LDA can not be altered. 
In contrast to the case, the connectivity is correctly altered when
we include the off-diagonal elements (or fully include the non-locality).

To plot the energy-band dispersion in the whole Brillouin zone (BZ),
we have to know non-local potential $\vxc_\bfk(\bfr,\bfr')$
at any $\bfk$ point in the BZ by an interpolation, 
where $\bfr$ and $\bfr'$ are within the primitive cell.
This interpolation is also needed for the offset-$\Gamma$ method in Sec.\ref{sec:kint}, 
and useful for calculating physical quantities which requires integrations in the BZ.
For the interpolation, we inevitably require real-space representation 
$\vxc(\bfr,\bfR'+\bfr')$, where $\bfR'$ is to specify origin of primitive cells.
If it is inverse Fourier transformed, we obtain $\vxc_\bfk(\bfr,\bfr')$ at any $\bfk$.
The nature that the MTOs are atom-centered and localized basis
enables us to make such an interpolation in FP-LMTO-QSGW
\cite{kotani_quasiparticle_2007}.

Even though FP-LMTO-QSGW have been successfully applied to many cases, e.g.,
\cite{chantis_ab_2006,lukashev:195202,kotani_re-examination_2009,christensen_electronic_2010,svane_quasiparticle_2010,huang_electronic_2013}, it still has problems. 
A main problem originates from the FP-LMTO method, 
the applicability of which is limited to the systems that can be described only by the MTOs.
Because of this fact, we have to fill empty regions with empty spheres (ESs) in, e.g., not closed packed systems and surfaces.
The effort of this procedure enforces us to repeat many 
calculations to check numerical convergence. Therefore, 
it was not easy to apply \QSGW\ to, e.g., surfaces. 
In addition, it is not so easy to enlarge 
basis set systematically as in the case of the LAPW method.
Furthermore, the interpolation of $\vxc_\bfk(\bfr,\bfr')$
was unstable in cases because we needed to use not well-localized MTOs 
(they contain damping factor $\propto \exp (-r/\kappa)$ 
where $\kappa^{-2} \sim$ 0.1 Ry. Thus the MTOs has long range). 
This required us to use a very complicated
interpolation procedure \cite{kotani_quasiparticle_2007}.

To overcome the problem in the FP-LMTO in DFT, we recently have given a new all-electron
full-potential first-principles method of the electronic structure calculations in
the GGA/LDA (one-body problem solver) \cite{kotani_fusion_2010,kotani_linearized_2013}.
This method is named the linearized augmented plane wave and muffin-tin
orbital method (the PMT method), which is a mixed basis method of augmented
waves, the APWs and the MTOs. Within our knowledge,
there is no other mixed-basis method of augmented waves.
We can use a procedure to set parameters of the MTOs almost automatically 
as given in Ref.\onlinecite{kotani_linearized_2013}.
(see Fig.1 and Table.I around in Ref.\onlinecite{kotani_linearized_2013}).
The important point is that  a serious difficulty 
in the FP-LMTO, how to set the parameters, is now overcomed. 
In the usual FP-LMTO,
we need to repeat many calculations to figure out reasonable 
parameters of the MTOs. In contrast, we can check convergence only by
changing number of APWs.
Based on the procedure, we find that 
the highly localized MTOs ($\kappa^{-2} = 1\sim 2$ (bohr)$^{-2}$) 
in combination with APWs whose cutoff energy $\sim$ 4Ry can 
give good convergence of total energy in the GGA/LDA \cite{kotani_linearized_2013};
we successfully obtained atomization energy for homo-nuclear diatomic molecules, 
which is converged more than chemical accuracy ($\sim$ 1 Kcal/mol).
Thus we can expand eigenfunctions with highly localized atom-centered MTOs and low energy APWs.

In this paper, we show how to implement the QSGW method in the PMT
method, that is, the PMT-QSGW method. 
After we explain the QSGW theory in Sec.~\ref{sec:qsgwth},
we explain the implementation of PMT-QSGW in Sec.~\ref{sec:impl}.
Especially, in Sec.~\ref{sec:kint}, we show new improvement to the
offset-$\Gamma$ method
to take into the anisotropy of the screened Coulomb interaction accurately;
in Sec.~\ref{sec:siginterp}, we explain the interpolation of
$\vxc_\bfk(\bfr,\bfr')$. 
Finally, in Sec.\ref{sec:numtest}, we show detailed 
numerical tests for the band gap (at $\Gamma$ point)
for GaAs and cubic SiO$_2$ ($\beta$-cristobalite).
We show how the \QSGW\ band gap can depends on the cutoff parameters.

\section{Theory of the \QSGW\ method}
\label{sec:qsgwth}
Here we summarize the \QSGW\ method. 
We treat the following many-body Hamiltonian for electronic system. 
With the field operators $\hat{\psi}_\sigma ({\bf r})$, spin index $\sigma$, 
external potential $\vext_{\sigma}({\bf r})$, 
and the Coulomb interaction $v(\bfr, \bfr')=\frac{e^2}{|\bfr-\bfr'|}$,
it is written as
\begin{eqnarray}
&&\hat{H}= \hHk + \hVee +\hVext, \label{eq:hami}\\
&&\hHk =     \sum_\sigma \int d {\bf r} 
        \hat{\psi}^{\dagger}_{\sigma}({\bf r})
        ( - \frac{\nabla^2}{2m} )
        \hat{\psi}_{\sigma}({\bf r}), \\
&&\hVext \!=\!\sum_{\sigma} \int d {\bf r} \vext_\sigma ({\bf r}) 
                              \hat{n}_\sigma ({\bf r} ), \label{eq:hvext}\\
&&\hVee \!\!\!=\!\frac{e^2}{2}\!\!
          \sum_{\sigma\sigma'}\! \int \!\!d^3\!r d^3\!r'\! v(\bfr,\!\bfr') 
           { \hat{\psi}^{\dagger}_{\sigma}\!({\bf r})
                   \hat{\psi}^{\dagger}_{\sigma'}\!({\bf r}')
                   \hat{\psi}_{\sigma'}\!({\bf r}')
                   \hat{\psi}_{\sigma}\!({\bf r}) }. 
\end{eqnarray}
Here, we omit classical electrostatic nucleus-nucleus
energy for simplicity. We explicitly show electron mass $m$ and charge
$e^2$ in the following formulas but with $\hbar=1$.
$\vext({\bf r})$ mainly contains those coming from nucleuses, in addition to
perturbation such as external magnetic fields. 
Hats on symbols mean the second quantized quantities
(for example, $\hVext$ and $\vext_\sigma(\bfr)$ 
mean the same physical quantities in different representations). 
In the followings, we omit the spin index $\sigma$
and often $\bfr$ for simplicity.

Let us consider how to obtain best one-body Hamiltonian $\H0$ 
to describe QPs for given $\hat{H}$.
If we have the self-energy $\Sigma(\bfr,\bfr',\omega)$ for given $\hat{H}$,
we can determine the QP energies and eigenfunctions as the solutions of
\begin{eqnarray}
H(\epsilon_i)  |\Psi_i (\bfr)\rangle = \epsilon_i |\Psi_i(\bfr)\rangle,
\label{eq:defqp}
\end{eqnarray}
at least near the Fermi energy, where 
the one-particle dynamical effective Hamiltonian $H(\omega)$ is
\begin{eqnarray}
H(\omega)= -\frac{\nabla^2}{2m} + \vext +\vh + \Sigma(\omega).
\label{eq:hdyn}
\end{eqnarray}
Here $\vext$ is the external potential from nucleus and $\vh$ is the Hartree potential. 
If $H(\omega)$ were $\omega$-independent and Hermitian, we could have directly identified
this as $\H0$. Apparently this is not true, however, based on the Landau-Silin's QP theory
(or the independent-particle picture), 
we can still expect that physical properties can be evaluated
with the use of the eigenvalues and eigenfunctions of the QPs $\{\epsilon_i,\Psi_i(\bfr)\}$.
This means a physical picture that primary excitations are specified by electrons 
or holes added to these orbitals, then 
they interact each other with a screened Coulomb interaction. 
A theoretical inconvenience is that the set of QPs is not a complete set
since $\Sigma(\omega)$ is energy-dependent non-Hermitian.
If we can use a static Hermitian one-body potential $\vxc(\bfr,\bfr')$ 
in place of $\Sigma(\omega)$,  such a problem does not occur.
Then the set $\{\Psi_i(\bfr)\}$ is an orthonormal complete set, and
physical quantities can be represented in the Fock space of the set.
In other words, we divide
the full many-body Hamiltonian $\hat{H}$ into 
$\hat{H}=\hH^0+ (\hat{H}-\hH^0)$, 
where the many-body contribution due to $(\hat{H}-\hH^0)$,
should not change the QPs given by $\hH^0$.

Following the above discussion, we need two methods 
to obtain such $\hH^0$ for given $\hat{H}$. These are
\begin{itemize}
\item[(i)] A method to calculate $\Sigma(\omega)$ (and also $\vh$)
for given division of $\hat{H}=\hH^0+ (\hat{H}-\hH^0)$.
\item[(ii)] 
A method to determine 
$\vxc$ as a good substitution of given $\Sigma(\omega)$.
\end{itemize}
If both methods (i) and (ii) are given, we can make a self-consistent cycle closed.
That is, we have the cycle 
$\hH^0 \rightarrow \{\vh,\Sigma(\omega)\} \rightarrow \hH^0 \rightarrow ...$.
This is repeated until converged. 
In QSGW, we use the $GW$ approximation for (i).
As for (ii), we use a mapping from $\Sigma(\omega)$
to a static Hermitian potential $\vxc(\bfr,\bfr')$ as
\begin{eqnarray}
\vxc &=& \frac{1}{2}\int_{-\infty}^{\infty} d \omega {\cal
 R}\left[\Sigma(\omega)\right]\delta(\omega-\H0) + {\rm c.c.} \nonumber \\
&=&\sum_{ij} |\Psi_i\rangle 
  \langle\Psi_i| \frac{{\cal R}\left[\Sigma(\ei)+\Sigma(\ej)\right]}{2} |\Psi_j\rangle
  \langle\Psi_j|,
\label{eq:vxceq}
\end{eqnarray}
where ${\cal R}[X]= \frac{X+X^\dagger}{2}$ means taking the Hermitian
part of $X$. 
\req{eq:vxceq} is given so as to
reproduce $\{\epsilon_i,\Psi_i(\bfr)\}$ satisfying \req{eq:defqp} 
as good as possible (this is not an unique choice \cite{kotani_quasiparticle_2007}). 
If necessary, we can derive \req{eq:vxceq}
from a minimization of the difference $G^{-1}-(G^0)^{-1}$, written as
${\rm Tr}\left[\left(G^{-1}-(G^0)^{-1}\right) \delta((G^0)^{-1}) \left(G^{-1}-(G^0)^{-1}\right)\right]+ {\rm c.c.}$ \cite{vans06}. 
The $GW$ approximation together with \req{eq:vxceq} makes 
a fundamental equation of QSGW.

Let us detail steps of the QSGW calculation.
We start from a trial one-particle static Hamiltonian written as
\begin{equation}
\H0 = \frac{-\nabla^2}{2m} + \veff(\bfr,\bfr').
\label{eq:defh0}
\end{equation}
This $\H0$ is just the initial condition for iteration cycle, and 
does not affects to the final result.
The $GW$ method is applied to the division $\hat{H}=\hH^0+ (\hat{H}-\hH^0)$.
Its steps are as follows, (I)-(V).
\begin{enumerate}
\item[(I)] We have the non-interacting Green's function $G^0$ for given $\H0$. It is
\begin{equation}
G^0\left( \bfr,\bfr',\omega \right) 
= \sum\limits_i {
\frac{\Psi_i(\bfr) \Psi_i^*(\bfr')}
     {{\omega -\ei \pm i \delta }}},
\label{eq:defg}
\end{equation}
where $\{\ei,\Psi_i\}$ are eigenvalues and eigenfunctions of $\H0$.
\item[(II)] Calculate the dynamical screened Coulomb interaction $W$ as
\begin{equation}
 W = \epsilon^{-1} v = \left(1-v \Pi\right)^{-1} v,
\label{eq:defw}
\end{equation}
where we use the proper polarization $\Pi=-iG^0\times{}G^0$.
\item[(III)] Calculate the self-energy $\Sigma(\bfr, \bfr'\!\!, \omega)$ as
\begin{equation} 
\label{self_energy}
\Sigma(\bfr, \bfr'\!\!, \omega)
\!=\!\frac{i}{2\pi}\!\!\int\!\!d\omega'
G^0(\bfr,\bfr'\!\!,\omega-\omega')W(\bfr,\bfr'\!\!,\omega')e^{-i\delta\omega'}\!\!.
\end{equation}
\item[(IV)]
Simultaneously, we can calculate $\vh$ for electron density from $G^0$.
Together with $\vext+\vh$, we have 
$H(\omega)=\frac{-\nabla^2}{2m} + \vext+\vh +\Sigma(\omega)$.
The conventional one-shot $GW$ evaluates its (usually diagonal) expectation values.
\item[(IV)] From $\Sigma(\bfr, \bfr'\!\!, \omega)$, we obtain $\vxc$
	    through \req{eq:vxceq}.

\item[(V)]  For $\vxc$, we obtain a new $\H0$ as
$\H0 = \frac{-\nabla^2}{2m} + \vext + \vh + \vxc$.
From this, we do again from (I). 
\end{enumerate}
We repeat these steps until converged.
In the procedure (I), we assume a non-interacting ground state 
by filling electrons up to the Fermi energy. 
The self-consistency ensures that this ground state 
is stable as for the $GW$ approximation.

We emphasize the ability of the non-locality 
of the one-particle potential $\veff(\bfr,\bfr')$ to describe QPs,
in comparison with the ability of the local potential used in GGA/LDA.
The non-locality may be classified to two kinds.
One is the onsite non-locality which
can be also partially described by $U$ of LDA+$U$.
We have to introduce such a onsite non-local potential for the
exchange-correlation term to enhance the size of 
orbital magnetic moment \cite{solovyev_1998}, because local potential
can not give the time-reversal symmetry.
The other is the offsite non-locality that is important to give the
difference of eigenvalues between bonding and anti-bonding orbitals.
For example, we can imagine a non-local potential which 
behaves as a projector to push down only eigenvalues of the bonding orbital.
A local potential can hardly give this effect.

The QSGW method can be justified from a view of the $GW\Gamma$ theory which
takes into account the vertex $\Gamma$; 
Ishii, Maebashi and Takada\cite{ishii2010} gave analyses for the effects 
of the vertex $\Gamma$ in the self-energy and in the polarization function.
They claims that the effect of $\Gamma$ is virtually cancelled out.
An illustration of their claim is for the renormalization factor $Z$ contained in $G$; 
in the calculation of $\Sigma=G\times W\times \Gamma$,
this $Z$ contained in $G$ is cancelled out by the vertex $\Gamma$,
which is reduced to be $1/Z$ at $\bfq,\omega \to 0$.
That is, we see the cancellation $Z \times 1/Z$ for the QP weights
in $G\times \Gamma$ \cite{kotani_quasiparticle_2007}.
This illustration is generalized by the Ward identity, and
they concluded that we should use $\Sigma=G_0\times W$ 
rather than $\Sigma=G\times W$ when we neglect vertex correction ($\Gamma=1$). 
Along the context of \QSGW, we interpret their theory as 
''If we have a good $\H0$ which nearly describes the QPs,
we can calculate good self-energy to determine the QPs 
by $\Sigma=G_0\times W$.''. 
Although they gave no discussion about how to obtain $\H0$, 
we think that QSGW is a possible candidate to determine such a $\H0$.
As for the polarization function, they also give a discussion 
not to use $\Pi=-iG \times G$ but to use $\Pi=-iG_0 \times G_0$ 
for the proper polarization. This is reasonable because $\Pi=-iG \times G$ contains
the QP electron-hole excitations with too small weight 
$Z_{\rm occupied} \times Z_{\rm unoccupied}$. 
This discussion for the proper polarization is consistent with the fact that first-principles
calculations of dielectric functions with $\Pi=-iG_0 \times G_0$
gives good agreements with experiments 
\cite{arnaud_local-field_2001} (and the agreements are improved by 
taking into account the two-body correlations in the Bethe-Salpeter equation).
On the other hand, numerical calculations by Bechstedt et
al\cite{Bechstedt97} 
showed that poorness of $\Pi=-iG \times G$ is corrected if we include the 
contribution of $\Gamma$ to $\Pi$. This is consistent with our claim here.
The discussions here gives a support
of QSGW rather than the full self-consistent $GW$ methods
\cite{holm_fully_1998,ku_band-gap_2002,stan_fully_2006,
PhysRevB.81.085103,PhysRevB.88.075105}.

Let us consider two effects which are missing in the QSGW method.
One is the effect not in the $GW$ method utilized in the method (i)
(or, almost equivalently, how to improve $W(\omega)$ in the step (II)).
QSGW, for example, tends to give a slightly larger band gap than
experimental one \cite{kotani_quasiparticle_2007}, which is traced back to
slightly strong $W(\omega)$ (slightly small
screening effect) in the RPA. Thus, we need better $W(\omega)$ beyond the RPA.
Along this line, some works are performed until now: $W(\omega)$ including
pair excitations \cite{shishkin_accurate_2007}; $W(\omega)$ 
including phonons \cite{botti_strong_2013}; $W(\omega)$ including
a vertex correction \cite{ishii2010}.
The other missing effects are, e.g., the contribution to the self-energy due
to the low energy excitations such as the magnetic fluctuations and phonons.
Note that QSGW gives QPs, where charge fluctuation is already 
taken into account in the RPA self-consistently. Thus we expect that
the main missing contribution comes from the low energy excitations.
If such contribution to the self-energy is taken into account, 
the QP dispersion near the Fermi energy can be 
deformed; kink-like structure (mass enhancement)
is added just near the Fermi energy \cite{deng_hallmark_2012} on top of the QP
dispersion of QSGW as long as the effects due to such fluctuations is not too large.
From the opposite point of view,
this means that QSGW describes overall feature of energy bands including the
Fermi surface except such mass enhancement near the Fermi energy.
Such low energy part of self-energy may be calculated with $W(\omega=0)$ 
(neglecting energy dependence) based on the many-body perturbation
theory,  although we need to avoid double counting problem 
of the Feynman diagrams intrinsic in the first-principles many-body
perturbation theory \cite{springer_first-principles_1998}.
Not so much research have been performed along this line,
in contrast to the first-principles method combined with the 
dynamical mean field theory \cite{RevModPhys.68.13}.

Let us discuss about the total energy in QSGW.
Formally, the total energy can be given by an adiabatic connection, usually
specified by a parameter $\lambda$ changing from zero through unity as
$\hat{H}^\lambda = \hat{H}^0  + (\lambda \hVee -\hVeff_\lambda)$; this
path starts from $\hat{H}^0$ at $\lambda=0$, and
ends with $\hat{H}$ at $\lambda=1$ 
(note that $\hat{H}^0$ and $\hVeff_{\lambda=1}$ are 
the second-quantized expressions of $\H0$ and $\veff$).  
Along the path,  $\hVeff_\lambda$ is supposed to be chosen so that QSGW applied to
$\hat{H}^\lambda$ gives $\hat{H}^0$ for any $\lambda$.
Then the total energy is given as
\begin{widetext}
\begin{eqnarray}
E = E^0 \!+\! \int_0^1 \! \! d\lambda \frac{d E^\lambda}{d \lambda} 
= E^0 
+ \int_0^1 \! d\lambda \langle0_{\lambda}| \hVee |0_{\lambda}\rangle
- \int_0^1 \! d\lambda \langle0_{\lambda}|\frac{\partial \hVeff_\lambda}{\partial \lambda} |0_{\lambda} \rangle,
\label{eq:etot1}
\end{eqnarray}
\end{widetext}
where $|0_{\lambda}\rangle$ is the ground state for $\hat{H}^\lambda$.
\req{eq:etot1} is an exact formula without approximation.
As the lowest
order approximation, we replace $|0_{\lambda}\rangle$ with
$|0_{\lambda=0}\rangle$. Then we have the Hartree-Fock energy
calculated from the eigenfunctions of $\H0$.
As a more accurate approximation, 
we evaluate $\langle 0_\lambda| \hVee |0_\lambda \rangle$ 
in the random phase approximation (RPA); we apply it
to $\hat{H}^0$ whose ground state is $|0 \rangle$, with the interaction
of $\lambda \hVee$.  This gives the polarization function
$\Pi (1-\lambda v \Pi)^{-1}$, where $\Pi(\omega)$ is the polarization
function of the non-interacting ground state $|0\rangle$.
Then we have the RPA total energy,
\begin{eqnarray}
&&\ERPA =  E_0^{\rm k} + E_0^{\rm ext} + E_0^{\rm H} + E_0^{\rm x} + E^{\rm c},\label{eq:erpa}\\
&&E^{\rm c} = \frac{-i}{2}{\rm Tr}[  \log(1-v \Pi) + v \Pi] \label{eq:ecrpa}.
\end{eqnarray}
The derivative of $\ERPA$ with respect to the number of occupation
for the orbital $\{\epsilon_i, \Psi_i\}$ gives
\begin{eqnarray}
&&\frac{\partial \ERPA}{\partial n_i} =  
\langle \Psi_i  |-\frac{\nabla^2}{2m} + \vext +\vh + \Sigma(\epsilon_i)
|\Psi_i \rangle. 
\label{eq:derpadn}
\end{eqnarray}
Since $\langle \Psi_i  | \Sigma(\epsilon_i) |\Psi_i \rangle$
are the diagonal elements of \req{eq:vxceq},
this change (derivative) of the RPA energy equals to the QP energy given by the QSGW.
In addition, the minimization of right-hand side of \req{eq:derpadn}
as a functional of $\Psi_i$ gives \req{eq:defqp} 
if we can neglect $\Psi_i$ contained in $\Sigma(\epsilon_i)$.
These show that the QSGW is related to the 'RPA' total energy.
We need caution to the meaning of the QP energy $\epsilon_i$.
It is not the change of the total energy for one electron added/removed,
but the derivative for occupancy. 
This is common to the case of the Koopman-Slater-Janak's theorem.
This is related to the localization-delocalization problem
\cite{cohen_insights_2008}, where we need to know how the eigenvalue 
$\epsilon_i$ changes as a function of fractional
occupancy. One must recognize that $\epsilon_i$ must be calculated
for the fractional occupancy, where we expect that $\epsilon_i$ 
changes relatively linearly, and be integrated 
with changing the occupancy \cite{bruneval_gw_2009,bruneval_ionization_2012} 
in order to calculate ionization energies and so on.  

Originally the QSGW is proposed to treat solids, however, we today have
requirement to treat molecules on surface for such problems like catalysis.
In the case of molecules (zero-dimensional systems), 
there are not only continuous eigenvalues but also discrete ones in $\H0$. 
Even in this case, \req{eq:defqp} is the equation to determine eigenstates of the system. 
However, it is not trivial whether we can 
extract the independent-particle (or the QP) picture in the manner of QSGW.
Only limited number of publications on the \QSGW\ applying to molecules
are available now \cite{bruneval_gw_2009,bruneval_ionization_2012}, 
and not so much have been clarified yet. 

\section{Implementation}
\label{sec:impl}

In Sec.~\ref{sec:ov}, we show overview of the method to perform
the $GW$ calculation. We made some improvements to the method in
Refs.\onlinecite{kotani_all-electron_2002,kotani_quasiparticle_2007},
where we take some ideas from another $GW$ implementation 
given by Friedrich, Bl\"ugel, and Schindlmayr
\cite{friedrich_efficient_2010}.

In Sec.~\ref{sec:kint}, we show new improvement to the offset-$\Gamma$ method, which is
in order to treat $\bfk \to 0$ divergence of integrand for the self-energy calculation.
This improvement can correctly capture anisotropy of the screened Coulomb interaction,
although the previous offset-$\Gamma$ method in FP-LMTO-QSGW \cite{kotani_quasiparticle_2007}
is dangerous to treat anisotropic systems.

In Sec.~\ref{sec:siginterp}, we explain the interpolation of 
$\vxc_\bfk(\bfr,\bfr')$. The interpolation procedure
is simplified in comparison with that used in FP-LMTO-QSGW.

\subsection{overview}
\label{sec:ov}
In the PMT method \cite{kotani_fusion_2010}, 
the valence eigenfunctions for given $\H0$ are represented
in the linear combinations of the Bloch summed MTOs
$\chi^{\bfk}_{\brl{j}}({\bf r})$ and the APWs $\chi^{\bfk}_\bfG (\bfr)$;
\begin{eqnarray}
\label{eq:lmtopsi}
\Psikn(\bfr) = \sum_{\brl{j}} z^{{\bfk}n}_{\brl{j}}
\chi^{\bfk}_{\brl{j}}({\bfr})+ \sum_{\bfG} z^{\bfk n}_{\bfG}
\chi^{\bfk}_\bfG(\bfr),
\label{eqeigen}
\end{eqnarray}
where we use indexes of wave vector $\bfk$, band index $n$, 
reciprocal lattice vector $\bfG$. The MTOs in the primitive cell are 
specified by index of MT site $\bfR$, angular momentum $L=(l,m)$, and
$j$ for radial functions. 
As for core eigenfunctions, we calculate them in the condition
that they are restricted within MTs.
Then we consider contributions of the cores only to the exchange part 
defined in \req{eq:sigx} in the followings. 
(In other words, we apply core1 treatment in 
Ref.\onlinecite{kotani_quasiparticle_2007} for all cores.)

In Ref.\onlinecite{kotani_fusion_2010}, we have tested variety
of basis sets of MTOs with APWs, whose numbers are specified by
the APW cutoff energy $\EMAX$. 
Then we show a simple and systematic procedure 
to choose the MTO basis sets in Ref.\onlinecite{kotani_linearized_2013}.
With the procedure, we can perform 
stable and accurate calculations. 
In the procedure, we use a large set of MTOs 
(two or three MTOs per $L$ for valence electrons) together with APWs 
with rather low cutoff energy, typically, $\sim$4 Ry.
Thanks to the APWs, we can include only highly localized MTOs.
For the damping factors $\propto \exp(- \kappa r)$ contained in MTOs,
we use $\kappa^2=$ 1.0 and 2.0 (bohr)$^{-2}$. 
In Ref.\onlinecite{kotani_linearized_2013}, we have shown that it is not necessary 
to optimize the $\kappa$ parameters when we use large enough $\EMAX$ ($\sim$4 Ry)
as shown in Fig.1 of Ref.\onlinecite{kotani_linearized_2013}.
Other parameters to specify MTOs are also fixed in a simple manner.
The smoothing radii of the smooth Hankel functions, which
are the envelope function of the MTOs, are set to be one half of the MT radii.
Thus the MTOs are chosen essentially automatically, and
the convergence is checked only by $\EMAX$.
In addition, we do not need to use ESs because APWs is substituted for the MTO basis of ESs.
We have shown that such basis set works well in practice 
to determine the atomization 
energies of homonuclear dimers from H$_2$ through Kr$_2$ with the
convergence of chemical accuracy $\sim$ 1 Kcal/mol or less
in the DF calculation in the PBE exchange correlation functional in a large supercell
\cite{kotani_fusion_2010}. Note that such supercell calculations are 
tough tests for augmented wave methods (FP-LAPW requires very
high $\EMAX$ because of small MT radius; 
it is not easy to apply FP-LMTO because of no way to fill ESs).
In comparison with methods only using the localized basis set
such as \texttt{Gaussian} in quantum chemistry,
the PMT method is advantageous in the point that it can describe
scattering states (higher than zero level) accurately.

At first, we re-expand $\Psikn(\bfr)$ in \req{eq:lmtopsi}  
as a sum of the augmentation parts in the MTs and the
PW parts in the interstitial region.
\begin{eqnarray}
\Psikn(\bfr)
= \sum_{\bfR u}  \alpha^{{\bfk}n}_{\bfR u} \varphi^{\bf k}_{\bfR u}({\bf r})
 + \sum_{\bf G}  \beta^{{\bfk}n}_{\bf G} P^{\bf k}_{\bf G}({\bf r}),
\label{def:psiexp}
\end{eqnarray}
where the interstitial plane wave (IPW) is defined as
\begin{eqnarray}
P^{\bf k}_{\bf G}({\bf r}) =
\begin{cases}
 0                           & \text{if {\bf r}} \in \text{any MT} \\
\exp (i ({\bf k+G})\cdot{\bf r}) & \text{otherwise}
\end{cases}
\end{eqnarray}
and $\varphi^{\bf k}_{R u}(\bfr)$ are Bloch sums of the atomic functions
$\varphi_{R u}(\bfr)$ defined within the MT at $R$,
\begin{eqnarray}
\varphi^{\bf k}_{R u}({\bf r}) &\equiv& \sum_{\bf T} \varphi_{R u}({\bf r-R-T}) \exp(i {\bf k\cdot{}T}).
\end{eqnarray}
{\bf T} and {\bf G} are lattice translation vectors in real and reciprocal
space, respectively. 

In the $GW$ calculation, we need not only the basis set for
eigenfunctions, but also the basis set to expand the product of eigenfunctions.
The basis is called as the mixed product basis (MPB) $\{M^{\bf k}_I({\bf r}) \}$
first introduced in Ref.\onlinecite{kotani_all-electron_2002}.
The MPB 
consists of the product basis (PB) within MTs \cite{aryasetiawan_product-basis_1994}
and the IPW in the interstitial region.
Since $\{M^{\bf k}_I({\bf r}) \}$ contains IPWs which are not orthogonal,
we define dual for $\{M^{\bf k}_I({\bf r}) \}$ as
\begin{eqnarray}
&& |\tilde{M}^{\bf k}_{I} \rangle \equiv \sum_{I'}
   |M^{\bf k}_{I'} \rangle (O^{\bf k})^{-1}_{I'I} \, , \\
&& O^{\bf k}_{I'I} = \langle M^{\bf k}_{I'} |  M^{\bf k}_I \rangle.
\end{eqnarray}
From $v_{IJ}^\bfk= \langle M^{\bf k}_{I} |v|  M^{\bf k}_J \rangle$,  
we calculate eigenfunction for the generalized eigenvalue problem defined by
$\sum_J (v_{IJ}^\bfk - v^\bfk_\mu O^{\bfk}_{IJ} ) w_{\mu J}^\bfk = 0$ where
$v_\mu(\bfk)$ are the eigenvalues of the Coulomb interaction matrix.
Then we have the Coulomb interaction represented by matrix elements as 
\begin{eqnarray}
v(\bfk)=\sum_{\mu} | E^{\bfk}_\mu \rangle {v_\mu(\bfk)} 
\langle E^{\bfk}_\mu |,
\label{eqvcoue}
\end{eqnarray}
where we define a new MPB 
$|E^{\bf k}_\mu({\bf r})\rangle=\sum_J |M^\bfk_J\rangle w^\bfk_{\mu J}$ 
which is orthonormal and is diagonal to the Coulomb interaction $v(\bfk)$. 
For the all-electron full-potential $GW$ approximation,
\req{eqvcoue} is introduced in Ref.\onlinecite{friedrich_efficient_2010}.
This corresponds to the representation in the plane wave expansion 
$v(\bfk+\bfG,\bfk+\bfG')=\frac{4 \pi \delta_{\bfG \bfG'}}{|\bfk+\bfG|^2}$.
$\mu=1$ corresponds to the largest eigenvalue of $v_{\mu}$, and 
$v_{\mu=1}$ is $\sim \frac{4 \pi e^2}{|\bfk|^2}$, which is related to 
the divergent term discussed in Sec.\ref{sec:kint}.

\begin{widetext}
With the definition of 
$\langle A| B\rangle =\int d^3r A^*(\bfr) B(\bfr)$,
the exchange part of $\Sigma(\omega)$ is written as
\begin{eqnarray}
\Sigma^{\rm x}_{nm}(\bfq)=
\langle \Psiqn|\Sigma_{\rm x} |\Psiqm \rangle
&&=-\sum^{\rm BZ}_{{\bf k}}  \sum^{\rm  occ}_{n'}
\langle \Psiqn| \Psi_{{\bf q-k}n'} E_\mu^{\bf k} \rangle
v_{\mu}({\bf k})
\langle E_\mu^\bfk \Psi_{{\bf q-k}n'} | \Psiqm \rangle.
\label{eq:sigx}
\end{eqnarray}

The screened Coulomb interaction $W(\omega)$ is calculated
through \req{eq:defw}, where the polarization function $\Pi(\omega)$
is written as
\begin{eqnarray}
\Pi_{\mu \nu}({\bf q},\omega)
&&=
\sum^{\rm BZ}_{\bfk} \sum^{\rm occ}_{n \ispone} \sum^{\rm unocc}_{n'\isptwo}
\frac{
\langle E^{\bf q}_\mu \Psikn |\Psi_{{\bf q+k}n'} \rangle
\langle \Psi_{{\bf q+k}n'}| \Psikn E^{\bf q}_\nu \rangle
}{\omega-(\varepsilon_{{\bf q+k} n'\isptwo}-\varepsilon_{\bfk n\ispone})+i \delta} \nonumber\\
&&+ \sum^{\rm BZ}_{\bfk} \sum^{\rm  unocc}_{n \ispone} \sum^{\rm occ}_{n'\isptwo}
\frac{
\langle E^{\bf q}_\mu \Psi_{{\bf k}n} |\Psi_{{\bf q+k}n'} \rangle
\langle \Psi_{{\bf q+k}n'}| \Psi_{{\bf k}n} E^{\bf q}_\nu \rangle
}{-\omega-(\varepsilon_{\bfk n\ispone}-\varepsilon_{{\bf q+k} n'\isptwo})+i \delta}.
\label{eq:polf0}
\end{eqnarray}
When time-reversal symmetry is assumed, $\Pi(\omega)$ can be simplified to read
\begin{eqnarray}
\Pi_{\mu \nu}({\bf q},\omega)
&&=\sum^{\rm BZ}_{{\bf k}}  \sum^{\rm  occ}_{n} \sum^{\rm  unocc}_{n'}
\langle E^{\bf q}_\mu \Psi_{{\bf k}n} |\Psi_{{\bf q+k}n'} \rangle
\langle \Psi_{{\bf q+k}n'}| \Psi_{{\bf k}n} E^{\bf q}_\nu \rangle \nonumber \\
&& \times
\left(\frac{1}{\omega-\varepsilon_{{\bf q+k}n'}+\varepsilon_{{\bf k}n}+i \delta}
-\frac{1}{\omega+\varepsilon_{{\bf q+k}n'}-\varepsilon_{{\bf k}n}-i \delta}\right). \label{dieele}
\label{eq:polf}
\end{eqnarray}
To evaluate \req{eq:polf0} or \req{eq:polf},
we first accumulate its imaginary parts (anti-Hermitian part) of $\Pi_{\mu \nu}(\bfq,\omega)$
along bins of histograms on the real axis $\omega$
with the tetrahedron technique \cite{rath_generalized_1975},
and then determines the real part via the Hilbert transformation.
The bins are dense near the Fermi energy and coarse at high energy as described in 
Ref.\onlinecite{kotani_quasiparticle_2007}.
This procedure is not only more efficient but also safer than methods
to calculate the real part directly. 
We also use the extended irreducible zone (EIBZ) 
symmetrization procedure described in Ref.\onlinecite{friedrich_efficient_2010}.

The correlation part of the screened Coulomb interaction $W^c(\omega)=W(\omega)-v$, 
which is calculated from $v$ and $\Pi(\omega)$ is given as
\begin{eqnarray}
W^{\rm c}(\bfk,\omega)=\sum_{\mu\nu} | E^{\bfk}_\mu \rangle {W^{\rm c}_{\mu\nu}(\bfk,\omega)} 
\langle E^{\bfk}_\mu |.
\end{eqnarray}
With this $W^{\rm c}(\bfk,\omega)$, we have the correlation part of the self-energy  as
\begin{eqnarray}\
\Sigma^{\rm c}_{n,n'}(\bfq,\omega)= \sum_{\bfk,m} \int_{-\infty}^\infty  \!d \omega' \sum_{\mu,\nu} 
\frac{\langle\Psiqn|\Psiqkm E^{\bfk}_\mu \rangle 
W^{\rm c}_{\mu\nu}(\bfk,\omega')\langle E^{\bfk}_\nu \Psiqkm
|\Psiqnp\rangle e^{-i \delta \omega'}}
{\omega-\omega'-\epsilon_{\bfq-\bfk m}\pm i \delta}.
\label{sigmann}
\end{eqnarray}
\end{widetext}
Here, we use $+i \delta$ for occupied states of
${\bfq\!-\!\bfk m}$, $-i \delta$ for unoccupied states.
In QSGW, 
we have to calculate Hermitian part of $\Sigma_{nn'}(\bfq,\epsilon_{\bfq n})$, 
in order to obtain $\vxc_\bfq$ via \req{eq:vxceq}.

There are two key points to handle the $GW$ procedure given above.
The first key point, given in Sec.\ref{sec:kint},
is the improved offset-$\Gamma$ method which treats
divergence of $W^{\rm c}(\bfk \to 0,\omega)$ in \req{sigmann}. 
For this purpose, we define non-divergent 
effective interaction $\overline{W^{\rm c}}(\bfk=0,\omega)$ instead of
${W^{\rm c}}(\bfk=0,\omega)$.
Then we can take simple discrete sum for both expressions of \req{eq:sigx} and \req{sigmann}.

The second point in Sec.\ref{sec:siginterp} is
how to make 
an interpolation to give $\vxc_\bfq$ at any $\bfq$ in the whole BZ,
from $\vxc_\bfq$ calculated only at limited numbers of $\bfq$ points.
This is required in the offset-$\Gamma$ method shown in Sec.\ref{sec:kint},
that is, we have to calculate eigenfunctions at some $\bfq$ points near $\bfq=0$.
For the interpolation, we expand the static
non-local potential $\vxc$ in \req{eq:vxceq} in the highly-localized
MTOs in the real space. Thus the MTOs are used for two purposes;
one is as the bases for the eigenfunctions, the other is as the bases to
expand $\vxc$. The interpolation procedure of $\vxc_\bfk(\bfr,\bfr')$
becomes stabilized and simplified rather than the
complicated interpolation procedure in 
Ref.\onlinecite{kotani_quasiparticle_2007}.
This is because we now use highly localized MTOs.
In the planewave-based \QSGW\ method by Hamann and Vanderbilt \cite{hamann09}, 
they expand $\vxc$ in the maximally localized Wannier functions instead of the MTOs.

In practical implementation, the LDA or GGA exchange-correlation
potential $V^{\rm xc}_{\rm LDA}$ 
is used as an assistance in order to generate core
eigenfunctions and also the radial functions within the MTs
(in this paper, we use subscript LDA even in the GGA. ``LDA/GGA'' means LDA or GGA).
The difference $\vxc-V^{\rm xc}_{\rm LDA}$ is used 
for the interpolation procedure in the BZ (explained in Sec.\ref{sec:siginterp}),
because this difference is numerically small 
as long as $V^{\rm xc}_{\rm LDA}$ is not so bad approximation.
These procedures with $V^{\rm xc}_{\rm LDA}$ give a slight dependence to
the final numerical results in practice as seen in Sec.\ref{sec:numtest},
although the results formally does not depend on the LDA/GGA exchange-correlation functions anymore.

\subsection{Improve offset-$\Gamma$ method}
\label{sec:kint}
The offset-$\Gamma$ method, originally 
invented for Ref.\onlinecite{kotani_all-electron_2002} by Kotani 
(described in Ref.\onlinecite{kotani_quasiparticle_2007}), 
was a key to perform accurate $GW$ calculation.
It is for integration of $\bfk$ 
in \req{eq:sigx} and \req{sigmann}, where
we have the integrands diverge at $\bfk \to 0$.
It worked well for highly symmetric systems, however,
it can be problematic to apply less symmetric systems, 
because anisotropic divergence of the integrands 
may not be treated accurately.
Here we show an improved offset-$\Gamma$ method, which treat
anisotropy of $W(\bfk,\omega)$ accurately.
In the followings, we use expression $W(\bfk)$
for simplicity (omit subscripts and $\omega$)
instead of $W_{\mu \nu}(\bfk,\omega)$, 
since we concern the $\bfk$ integral.


Let us give a formula to calculate 
$\int_{\rm BZ} f(\bfk) d^3k$ by discrete sum on $\bfk$-mesh,
where $f(\bfk)=G(\bfq-\bfk) \times W(\bfk)$.
As the $\bfk$-mesh, we use
\begin{eqnarray}
{\bf k}(i_1,i_2,i_3) &=& 2 \pi (\frac{i_1}{N_1} {\bf b}_1 
+ \frac{i_2}{N_2} {\bf b}_2 + \frac{i_3}{N_3} {\bf b}_3),
\label{kmesh}
\nonumber
\end{eqnarray}
where ${\bf b}_1,{\bf b}_2$ and ${\bf b}_3$ are the primitive reciprocal
vectors (the same as the Eq.(47) in Ref.\onlinecite{kotani_quasiparticle_2007}). 
The 1st BZ is divided into $N=N_1 \times N_2 \times N_3$
microcells ($i_1=0,1,... N_1-1$. Also the same for $i_2$, and $i_3$.).
The microcell including the $\Gamma$ point is called as the
$\Gamma$ cell \cite{freysoldt_dielectric_2007}.
Main problem is how to evaluate the contribution from the $\Gamma$ cell.
The divergent part of $f(\bfk)$ behaves $\approx$ (analytic function of $\bfk$) 
$/(\bfk^{\rm T}{\bf L}\bfk)$, 
where $\bfk^{\rm T}$ means the transpose of $\bfk$, ${\bf L}$
is an $3\times3$ Hermitian matrix \cite{friedrich_efficient_2010}. 
We neglect an odd part of $\bfk$ in the above (analytic function of $\bfk$)
because it gives no contribution to the integral around $\bfk=0$.
Thus it is enough to consider integral for $f(\bfk)$ whose divergent parts behaves
$f(\bfk)=\sum_L \frac{f_L Y_L(\widehat{\bfk})}{|\bfk|^2}$ at $\bfk \to 0$, 
where $l$ of $L\equiv(l,m)$ are restricted to be even number.
We evaluate the integral by a formula
\begin{eqnarray}
\int_{\rm BZ} f(\bfk) d^3k \approx \frac{1}{N}\sum^{\bfk \ne 0} f(\bfk)
 + \sum_L f_L w_L + \frac{1}{N} \tilde{f},
\label{eq:bzint0}
\end{eqnarray}
which is introduced in Ref.\onlinecite{freysoldt_dielectric_2007}. 
Here weights $w_L$ are determined in a manner as follows,
so as to take into account contributions of divergent part of $f(\bfk)$
at $\bfk \to 0$ in the $\Gamma$ cell.
$\tilde{f}$ is the constant part of $f(\bfk)$ at $\bfk \to 0$.

To determine $w_L$, 
we can use the following procedure instead of that given 
in Ref.\onlinecite{freysoldt_dielectric_2007}.
We first introduce auxiliary functions  
\begin{eqnarray}
F_L(\bfk) =\sum_{\bfG} \frac{\exp(- \alpha
 |\bfk-\bfG|^2)Y_L(\widehat{\bfk-\bfG})}{|\bfk-\bfG|^2}.
\end{eqnarray}
This is a generalization of an auxiliary function used in the 
offset-$\Gamma$ method (then we only used $F_{00}$ \cite{kotani_quasiparticle_2007}).
We usually take $\alpha \to 0$ limit, or small enough $\alpha$ instead.
Let us apply \req{eq:bzint0} to $F_L(\bfk)$.
Then we can evaluate the left hand side of \req{eq:bzint0}
exactly (the exact values are zero except $L=(0,0)$). 
On the other hand, the first term and the
third term in the right-hand side of \req{eq:bzint0}
can be evaluated numerically.
In addition, we know that $f_{L'}$ for $F_L(\bfk)$ is unity for $L'=L$,
and zero otherwise.
Thus we can determine $w_L$ in \req{eq:bzint0}
so that \req{eq:bzint0} is exactly satisfied for $F_L(\bfk)$ for any $L$.


Let us apply \req{eq:bzint0} to $f(\bfk)=G(\bfq-\bfk) \times W(\bfk)$.
Then we make an approximation taking only the most divergent term 
in $W(\bfk)$ in addition to its analytic part. That is, we use
\begin{eqnarray}
W_{\mu \nu}(\bfk) \sim \widetilde{W}_{\mu \nu}(\bfzero) +
 \frac{4 \pi}{\bfk^{\rm T} {\bf L} \bfk}
 \delta_{1\mu}\delta_{1\nu}
\label{eq:wnear0}
\end{eqnarray}
at $\bfk \to 0$. 
$\widetilde{W}_{\mu \nu}(\bfzero)=0$ for $\mu=1$ or $\nu=1$. 
See Eq.(36) in Ref.\onlinecite{friedrich_efficient_2010} to know what is
neglected in the approximation of \req{eq:wnear0}.

Then we finally obtain
\begin{eqnarray}
\int_{\rm BZ} d^3k G(\bfq-\bfk) W(\bfk) 
\approx \overline{\sum G(\bfq-\bfk) W(\bfk)},\label{eq:bzintbar}
\end{eqnarray}
where its right-hand side is defined as
\begin{eqnarray}
&&\overline{\sum G(\bfq-\bfk) W(\bfk)} \nonumber \\
&&\equiv 
\frac{1}{N} \sum_{\bfk \ne 0} G(\bfq-\bfk) W(\bfk) + \frac{1}{N} G(\bfq)\overline{W}(\bfzero),\\
&&\overline{W}(\bfzero)\equiv N\sum w_L W_L + \widetilde{W}(\bfzero).
\end{eqnarray}
Here $\overline{W}(\bfzero)$ can be taken as an
averaged $W$ in the $\Gamma$ cell.
With this $\overline{W}(\bfzero)$, we can 
evaluate integrals just by sum on discrete $\bfk$-mesh.
When the matrix ${\bf L}$ is given 
(a method to calculate ${\bf L}$ is given in the next paragraph),
the non-analytic (but non-divergent) function 
${\bfk^{\rm T}{\bf L}\bfk}/|\bfk^2|$ 
is expanded in the spherical harmonics. Then
$W_L$ is calculated for the given ${\bf L}$ in the manner of Ref.\onlinecite{friedrich_efficient_2010}. 
We can evaluate the accuracy of integrals with discrete $\bfk$-mesh
in combination with the approximation \req{eq:wnear0} 
by calculations with changing the size of the $\bfk$-mesh.




The remaining problem is how to calculate
the matrix ${\bf L}$ in \req{eq:wnear0}.
There are two possible ways to determine it.
One is the $\bfk \cdot \bfp$ method (perturbation) used in \cite{friedrich_efficient_2010}, 
the other is numerical method to determine them by calculations at some
$\bfk$ points near $\bfk=0$. Here we use the latter method.
Because of the point-group symmetry of the system, ${\bf L}$ can be expressed by the
linear combination of invariant tensors $\mu_{ij}^g$ for the symmetry of the unit cell;
\begin{eqnarray}
L_{ij}(\omega)=  \sum_{g=1}^{N_g} a_g(\omega) \mu_{ij}^g,
\end{eqnarray}
where $g$ is the index of invariant tensor. 
The number of $g$, $N_g$, can be
from one (cubic symmetry) through six (no symmetry).
It is possible to determine coefficient $a_g(\omega)$
from the dielectric functions ${\hat{\bfk}_{0i}}^{\rm T}{\bf L}\hat{\bfk}_{0i}$
calculated at $\{\bfk_{0i}\}$ points around
$\bfk=0$, where $\{\bfk_{0i}; i=1,N_g\}$ is a set of the offset-$\Gamma$ points.
The offset-$\Gamma$ points are chosen so that conversion matrix
from $\hat{\bfk}_{0i}^{\rm T}{\bf L}(\omega)\hat{\bfk^{0i}}$ to $a_g(\omega)$
should not numerically degenerated. The length $|\bfk^{0i}|$
can be chosen to be small enough, but avoiding numerical error
%
as the average of $W(\bfk)$ in the $\Gamma$ cell.
The improved offset-$\Gamma$ method shown here can be applicable even to metal cases, as long as
$\hat{\bfk}_{0i}^{\rm T}{\bf L}(\omega)\hat{\bfk_{0i}}$ contains the contribution
due to intraband transition.


\subsection{Interpolation of the self-energy in the Brillouin zone}
\label{sec:siginterp}
Here we show 
an interpolation procedure to give
$\vxc_{\bfk}$ at any $\bfk$,
from $\vxc_{\bfk}$ calculated only at the regular mesh points $\bfk(i_1,i_2,i_3)$.
This interpolation is used for the offset-$\Gamma$ method that
requires $W(\omega)$ at $\{\bfk_{0i}\}$;
to calculate these $W(\omega)$, we need 
eigenfunctions and eigenvalues not only 
at the regular mesh points $\bfk(i_1,i_2,i_3)$,
but also at $\bfk(i_1,i_2,i_3)+\bfk_{0i}$. 
This interpolation is also useful to plot 
energy bands, thus to obtain effective mass and so on. 
A key point of the interpolation is that $\vxc$ 
is expanded in real space in the highly localized MTOs as follows.

At the end of the step of (IV) in Sec.\ref{sec:qsgwth}, we obtain the matrix elements
$\langle \Psikn | \Dvxc_\bfk | \Psikm \rangle$ on the regular mesh
points of $\bfk$, where
$\Dvxc_\bfk=\vxc_\bfk-V^{\rm xc,LDA}_\bfk$.
Then it is converted to the representation in the APW and MTO bases as
\begin{eqnarray}
\langle \chi^{\bfk}_a| \Dvxc_\bfk | \chi^{\bfk}_b \rangle
= \sum_{n,m} \left(z^{-1}\right)^*_{an}\langle \Psikn |\Dvxc_\bfk |
\Psikm \rangle z_{bm}^{-1}, \nonumber \\
\label{eqvxcchi}
\end{eqnarray}
where we use simplified basis index $a$, which is the index to specify a basis 
($\brl{j}$ for MTO or ${\bfG}$ for APW).
Thus $\chi^{\bfk}_a$ denotes the APWs or MTOs in \req{eqeigen};
$z_{na}$ ($\bfk$ is omitted for simplicity)
means the coefficients of the eigenfunctions at $\bfk$, that is,
$z^{{\bfk}n}_{\brl{j}}$ and $z^{\bfk n}_{\bfG}$ in \req{eqeigen} together.
This $z_{an}$ is identified as a conversion matrix which connect
eigenfunctions (band index $n$) and the APW and MTO bases (basis index $a$).

To obtain real space representation, we need a representation
expanded in the basis that consist of the Bloch summed localized orbitals, 
which are periodic for $\bfk$ in the BZ. However, this is not the case for the APWs in
\req{eqvxcchi}. To overcome this problem, we make an approximation 
that we only take the matrix elements 
related to the MTOs, that is, the elements 
$\langle \chi^{\bfk}_a| \Dvxc_\bfk | \chi^{\bfk}_b \rangle$
where $a$ and $b$ specify MTOs.
The part related to APWs are not thrown away but projected onto the basis of MTOs.
This approximation can be reasonable as long as main part of $\Dvxc$ can be
well expanded in the MTOs,
although we need numerical tests to confirm accuracy as shown in Sec.\ref{sec:numtest}.
Then we obtain a real-space representation of 
$\Dvxc$ expanded in the MTOs from the MTO part of 
$\langle \chi^{\bfk}_a| \Dvxc_\bfk | \chi^{\bfk}_b \rangle$ by the Fourier transformation.
Then we can have interpolated one by the inverse Fourier transformation from it for any $\bfk$.
Since we use highly localized MTOs, 
this interpolation procedure is more stable 
than the previous one in the FP-LMTO-QSGW \cite{kotani_quasiparticle_2007}.
A complicated interpolation procedure 
given in Sec.II-G in Ref.\onlinecite{kotani_quasiparticle_2007} is not necessary
anymore.

To reduce computational time,
we calculate $\langle \Psikn |\Dvxc_\bfk | \Psikm \rangle$ only up to the states whose
eigenvalues are less than $\EMAXS$. 
Then the higher energy parts of matrix elements is assumed 
to be diagonal, where their values are given by a constant, 
an average of calculated diagonal elements.

\section{Numerical test}
\label{sec:numtest}
\begin{table}[htb]
\begin{center}
\caption{\label{tab:mtoset}
Used MTOs for GaAs and SiO2c ($\beta$-cristobalite). These are specified
by the principle quantum numbers and angular momentums.
The MTO's envelope functions are the smooth Hankel functions,
which are specified by two parameters, 
the damping factor $\kappa$ and the smoothing radius $R_{\rm sm}$.
We set the parameters in the manner of Ref.\onlinecite{kotani_fusion_2010}.
$R_{\rm sm}$ is given to be one half of the MT radius $R_{\rm MT}$, which is shown in 
the unit of bohr radius. Empty spheres (ESs) are located 
in the middle of the interstitial region
(two ESs per primitive cell in both of GaAs and SiO2c).
ESs are used only cases specified by ``vwn,es'' and ``pbe,es'' 
in \ref{tab:bandgaplda} and \ref{tab:bandgapqsgw}.
Unit of $\kappa^2$ is in (bohr)$^{-2}$}
\begin{tabular}{lcccc|l}
\toprule
   & valence & $R_{\rm MT}$ \\
\colrule
GaAs\\
   Ga  & 3d(lo), 4s4p4d4f($\kappa^2\!=\!1.0$), 4s4p4d($\kappa^2\!=\!2.0$)  & 2.19 \\
   As  & 3d(lo), 4s4p4d4f($\kappa^2\!=\!1.0$), 4s4p4d($\kappa^2\!=\!2.0$)  & 2.30 \\
   (ES)& 1s2p3d($\kappa^2\!=\!1.0$), 1s2p($\kappa^2\!=\!2.0$) & 2.80 \\
\colrule
\multicolumn{3}{l}{SiO2c (two Si and four O in a primitive cell)}   \\
   Si  & 4s4p4d4f($\kappa^2\!=\!1.0$), 4s4p4d($\kappa^2\!=\!2.0$) & 2.19 \\
   O   & 2s3p4d($\kappa^2\!=\!1.0$), 2s3p4d($\kappa^2\!=\!2.0$) & 2.30\\
   (ES)& 1s2p3d4f($\kappa^2\!=\!1.0$) & 2.80 \\
\botrule
\end{tabular}
\end{center}
\end{table}
Here we show results of test calculations for PMT-QSGW
applied to two examples, GaAs and the cubic SiO$_2$ ($\beta$-cristobalite,
denoted as SiO2c hereafter). The latter has large interstitial
regions; it has the same structure of Si but
oxygen atoms are located in the middle of Si-Si bonds.
We use lattice constants 5.653 \AA\ for GaAs, and 7.165 \AA\ for SiO2c.
We perform calculations with different settings
in order to show the convergence properties of the band gaps.
We use the simple and systematic procedure 
to determine sets of MTOs and APWs, as is  
explained after \req{eqeigen}.
We use the MTOs shown in Table.\ref{tab:mtoset}. 
As for Ga(3d) and As(3d), we use the local orbitals \cite{localorbital}.

\begin{table}[tb]
\begin{center}
\caption{\label{tab:bandgaplda}
Band gap in LDA/GGA to check the convergence on the basis set. 
For sets of MTOs shown in Table.\ref{tab:mtoset},
we tabulate the calculated band gaps for $\EMAX$.
Note ``vwn,es'' and ``pbe,es'' means with ESs. 
We can see band gaps converge well with small number of APWs;
this is consistent with the case of atomization energies \cite{kotani_linearized_2013}.
In the GGA case with ESs, 
convergence behavior becomes a little unstable (not converged for 6.0 Ry for SiO2c),
because of numerical instability of linear-dependency.
$n_{\rm APW}$ means number of APWs at $\bfk=0$.
Number of MTOs without ESs are 60 for GaAs and 168 for SiO2c.}
\begin{tabular}{cccccc}
\toprule
\multicolumn{6}{c}{band gap (eV) in LDA/GGA}   \\
\colrule
$\EMAX$(Ry)  & vwn  & vwn,es & pbe  & pbe,es &  $n_{\rm APW}$ \\
\colrule
GaAs \\
0.0 & 0.425 & 0.308 & 0.665 & 0.541 & 0\\
1.0 & 0.322 & 0.295 & 0.558 & 0.528 & 1\\
2.0 & 0.294 & 0.294 & 0.526 & 0.529 & 15\\
3.0 & 0.294 & 0.294 & 0.528 & 0.532 & 27\\
4.0 & 0.294 & 0.294 & 0.530 & 0.535 & 51\\
5.0 & 0.294 & 0.294 & 0.530 & 0.536 & 59\\
6.0 & 0.294 & 0.294 & 0.530 & 0.538 & 65 \\
\colrule
SiO2c\\
0.0 & 8.560 & 6.131 & 8.592 & 6.186 & 0 \\
1.0 & 5.406 & 5.434 & 5.663 & 5.563 & 15\\
2.0 & 5.437 & 5.445 & 5.670 & 5.622 & 27\\
3.0 & 5.442 & 5.445 & 5.665 & 5.652 & 59\\
4.0 & 5.444 & 5.446 & 5.665 & 5.668 & 65\\
5.0 & 5.446 & 5.447 & 5.668 & 5.658 & 113 \\
6.0 & 5.446 & 5.445 & 5.669 & ---   & 169 \\
\botrule
\end{tabular}
\end{center}
\end{table}
In advance to show the band gaps calculated in QSGW, let us show those
in LDA/GGA in Table.\ref{tab:bandgaplda}. 
We can check the convergence behavior by changing the APW cutoff energy $\EMAX$. 
For the functional of LDA, we use the VWN exchange-correlation functional
\cite{vwn}; for GGA, we employ PBE \cite{pbe};
'vwn,es' and 'pbe,es' mean cases that ESs are included.
The convergence behavior is satisfactory,
as was in the case of total energy for homo-nuclear dimers 
\cite{kotani_linearized_2013}.
We see better convergence behavior as for $\EMAX$ for 'vwn,es' and
'pbe,es' than 'vwn' and 'pbe', since we have larger number of basis.
For example,'vwn,es' for GaAs
shows 0.295 eV for $\EMAX=$1 Ry is essentially the same as the
converged value of 0.294 eV, while 'vwn' requires $\EMAX \gtrsim$ 2 Ry
to have similar convergence.
For SiO2c, the convergence is a little slower because SiO2c has large
interstitial region, e.g., the band gap 5.437 eV for 'vwn' at $\EMAX\gtrsim$2 Ry
shows $\sim$0.01 eV difference from converged value of 5.445 eV 
(we took the case of 'vwn,es' at $\EMAX$=6Ry).
Within this small error, we can determine the band gap even without ESs.
This confirms our expectation
that missing part of the Hilbert space spanned by highly localized MTOs 
(large damping factors $\kappa^2=$ 1.0 and 2.0 (bohr)$^{-2}$) 
is complemented by the APWs with such very low $\EMAX$.
The wave number of the cutoff corresponds to distance between nearest-neighbor atoms.
We saw a little instability (we need many iterations) in the calculations  
when we use $\EMAX\gtrsim$ 5 Ry in the case of 'pbe,es',
since GGA requires better numerical accuracy to calculate derivative of density.
This is because of the overcompleteness problem of the basis set, that is,
we lose linear-independency of basis functions for large $\EMAX$.
We conclude that Table.\ref{tab:bandgaplda} 
gives a satisfactory convergence behavior within this limitation.

\begin{table}[htb]
\begin{center}
\caption{\label{tab:pbset}
The product basis (PB) within MTs are constructed from the products of atomic basis.
After all the products are generated, remove linearly-dependent ones
with the use of the overlap matrix of the products. See
 Ref.\onlinecite{kotani_quasiparticle_2007} in detail. 
$l_{\rm cut}$ means the allowed maximum $l$ of the PB. 
$n_{\rm PB}$ shows the total number of PB in each MT.}
\begin{tabular}{llccccc}
\toprule
      & &products  & $l_{\rm cut}$ & tol & $n_{\rm PB}$ \\
\colrule
GaAs \\
\colrule
PB0 &Ga & $\phi(4s,4p,3d,4d) \times \phi(4s,4p,3d,4d,4f)$ & 4 & $10^{-3}$ & 97 \\
    &As & $\phi(4s,4p,3d,4d) \times \phi(4s,4p,3d,4d,4f)$ & 4 & $10^{-3}$ & 106 \\
\colrule
PB0$t$ &Ga &  PB0 & 4 & $10^{-5}$ & 119\\
    &As &  PB0 & 4 & $10^{-5}$ & 126\\
\colrule
PB0$l$ &Ga &  PB0 & 6 & $10^{-3}$ & 119\\
    &As &  PB0 & 6 & $10^{-3}$ & 128\\
\colrule

PB1 &\begin{tabular}{@{}l@{}} Ga\\\ \\\end{tabular}  &
         \begin{tabular}{@{}l@{}}
         $\phi(4s,4p,3d,4d) \times \phi(4s,4p,3d,4d,4f)$ \\
         $\phi(4s,4p,3d,4d) \times \phidot(4s,4p,3d,4d,4f)$
          \end{tabular}                          & 4 & $10^{-3}$ & 115\\
    &\begin{tabular}{@{}l@{}}    As\\\ \\\end{tabular}  &
         \begin{tabular}{@{}l@{}}
         $\phi(4s,4p,3d,4d) \times \phi(4s,4p,3d,4d,4f)$\\
         $\phi(4s,4p,3d,4d) \times \phidot(4s,4p,3d,4d,4f)$ 
         \end{tabular}                       & 4 & $10^{-3}$ & 115\\
  & (ES) & $\phi(1s,2p,3d) \times \phi(1s,2p,3d,4f)$     & 2      & $10^{-3}$ & 22 \\
\colrule
PB1$l$ &Ga &  PB1 & 6 & $10^{-5}$ & 175\\
    &As &  PB1 & 6 & $10^{-5}$ & 178\\
\colrule
SiO2c \\
\colrule
PB0 &Si & $\phi(3s,3p,3d) \times \phi(3s,3p,3d,4f)$  & 4 & $10^{-3}$ & 75 \\
    &O  & $\phi(2s,2p,3d) \times \phi(2s,2p,3d,4f)$  & 4 & $10^{-3}$ & 67 \\
\colrule
PB0s &Si & PB0  & 4 & $10^{-3}$ & 76 \\
    &O & PB0  & 2 & $10^{-3}$ & 31 \\
\colrule

PB1 &\begin{tabular}{@{}l@{}} Si\\\ \\\end{tabular}  &
         $\begin{tabular}{@{}l@{}}
         $\phi(3s,3p,3d) \times \phi(3s,3p,3d,4f)$\\
         $\phi(3s,3p,3d) \times \phidot(3s,3p,3d,4f)$
          \end{tabular}$                              & 4 & $10^{-3}$ & 76 \\
& \begin{tabular}{@{}l@{}}    O\\\ \\\end{tabular}  &
         $\begin{tabular}{@{}l@{}}
         $\phi(2s,2p,3d) \times \phi(2s,2p,3d,4f)$\\
         $\phi(2s,2p,3d) \times \phidot(2s,2p,3d,4f)$ 
         \end{tabular}$                              & 2 & $10^{-3}$ & 31\\
  &(ES) & $\phi(1s,2p,3d) \times \phi(1s,2p,3d,4f)$     & 2      & $10^{-3}$ & 22 \\
\colrule
PB1$l$ &Si & PB1 & 4 & $10^{-5}$ & 76 \\
    &O  & PB1 & 4 & $10^{-5}$ & 70 \\
\botrule
\end{tabular}
\end{center}
\end{table}

\begin{table*}[tb]
\caption{\label{tab:bandgapqsgw}
Band gap (at $\Gamma$) for GaAs 
and cubic SiO2c ($\beta$-cristobalite SiO$_2$) in the PMT-QSGW method
in different cutoffs/settings. Number of used $\bfk$ points in the 1st BZ
is $4\times4\times 4$ for GaAs, and $2\times 2 \times 2$ for SiO2c. 
No spin-orbit coupling. 
The first line named as REF is treated as a standard to compare others in this table.
Empty columns mean the default settings of REF.
The column $|{\bf q +G}|^{\Psi,W}_{\rm Max}$ shows used 
$|{\bf q +G}|^{\Psi}_{\rm Max}$ and $|{\bf q +G}|^{W}_{\rm Max}$.
Lines marked by ``*'' show best-efforts values(largest bases). 
Lines marked by ``*'' is the case used for Fig.\ref{fig:nkconv}.}
\begin{tabular}{cc}
\begin{tabular}[t]{lcccc|l}
\toprule
GaAs \\
   XC  & $|{\bf q +G}|^{\Psi,W}_{\rm Max}$ &  PB  & $\EMAXS$ & $\EMAX$ & Band gap \\
     & (1/bohr) &   &  (Ry) &  (Ry) &  (eV) \\
\colrule
\colrule
REF:  & & & & & \\
  vwn  &  4.0, 3.0 & PB1 & all & 3.0   &  1.939 \\
\colrule
\colrule
       &  6.0, 4.0 &      &     &       &  1.939  \\
       &  3.5, 3.0 &      &     &       &  1.940  \\
       &  3.0, 2.5 &      &     &       &  1.934  \\
\colrule
       &           & PB0  &     &       &  1.956  \\
       &           & PB0$t$  &     &       &  1.938  \\
       &           & PB0$l$  &     &       &  1.967  \\
       &           & PB1$l$  &     &       &  1.946  \\
\colrule
       &           &      &     & 2.0   &  1.931 \\
       &           &      &     & 4.0   &  1.950 \\
       &           &      &     & 5.0   &  1.959 \\
       &           &      &     & 6.0   &  1.969 \\
\colrule
       &           &      & 3.0 &       &  1.942 \ ** \\ 
       &           &      & 3.0 & 6.0   &  1.980 \\
\colrule
vwn,es &           &      &     &       &  1.945 \\
vwn,es &           &      &     & 6.0   &  1.982 \ \ * \\
vwn,es &           &      & 3.0 &       &  1.903 \\
vwn,es &           &      & 3.0 & 6.0   &  1.940 \\
\colrule
pbe    &           &      &      &  2.0    &  1.973 \\
pbe    &           &      &      &         &  1.981 \\
pbe    &           &      &      &  4.0    &  1.992 \\
pbe    &           &      &      &  5.0    &  2.001 \\
pbe    &           &      &      &  6.0    &  2.010 \\
pbe,es &           &      &      &         &  1.969 \\
pbe,es &           &      & 3.0  &         &  1.940  \\ 
pbe,es &           &      &      &  6.0    &  2.002 \ * \\
\botrule
\end{tabular}
\ \ \ & 
\begin{tabular}[t]{lcccc|l}
\toprule
SiO2c \\
   XC   & $|{\bf q +G}|^{\Psi,W}_{\rm Max}$ &  PB  & $\EMAXS$ & $\EMAX$ & Band gap \\
     & (1/bohr) &   &  (Ry) &  (Ry) &  (eV) \\
\colrule
\colrule
REF:  & & & & & \\
  vwn  &  4.0, 3.0 & PB1 & all & 3.0   &  11.16\\
\colrule
\colrule
       &  8.0, 6.0 &      &     &       &  11.28  \\
       &  6.0, 4.0 &      &     &       &  11.27  \\
       &  3.5, 3.0 &      &     &       &  11.15  \\
       &  3.0, 2.5 &      &     &       &  10.76  \\
\colrule
       &           & PB0  &     &       &  11.20  \\
       &           & PB0s  &     &       &  11.17  \\
       &           & PB1$l$  &     &       &  11.19  \\
       &           & PB1$l$  &     & 6.0   &  11.21  \\
\colrule
       &           &      &     & 6.0   &  11.18  \\
\colrule
       &           &      & 3.0 &       &  10.38 \  ** \\
       &           &      & 6.0 &       &  10.78 \\
       &           &      & 9.0 &       &  10.99 \\
\colrule
vwn,es &           &      &     &        &  10.49 \\
vwn,es &           &      &     & 4.0    &  10.47 \\
vwn,es &           &      &     & 6.0    &  10.41  \ * \\
vwn,es &           &      & 3.0 &        &  10.09 \\
\colrule
pbe    &           &      &      &         &  11.31  \\
pbe    &           &      &      &  6.0    &  11.33  \\
pbe,es &           &      &      &         &  10.57 \\
pbe,es &           &      &      &  5.0    &  10.54  \ * \\
\botrule
\end{tabular}
\end{tabular}
\end{table*}

Let us summarize settings (and parameters) to perform the PMT-QSGW
calculations.
These can be classified into followings;
\begin{enumerate}
\item[(A)] IPW cutoff $|{\bf q +G}|^\Psi_{\rm Max}$ to give allowed 
           $P^{\bf q}_{\bf G}({\bf r})$ in the expansion of eigenfunctions \req{def:psiexp}.
\item[(B)] Settings of the mixed product basis.
           We have parameters to specify product basis (PB) within MTs.
           The IPWs belonging to the mixed product basis 
	   is given by the cutoff $|{\bf q +G}|^W_{\rm Max}$.
           The sets of PB are shown in Table \ref{tab:pbset}.
\item[(C)] Cutoff energy for self-energy. 
	   As we explained in Sec.\ref{sec:siginterp},
	   we calculate $\langle i|\Dvxc |j\rangle$
	   only for $\epsilon_i \le \EMAXS$ and $\epsilon_j \le \EMAXS$, where $\EMAXS$
	   is measured from the top of valence. See the bottom of Sec.\ref{sec:siginterp}.
\item[(D)] Energy-axis parameters for GW. \\
	   These are used to accumulate imaginary part of $W(\omega)$.
	   See the explanation around \req{eq:polf}. 
	   We use an energy mesh (bin width);
	   the bin width is 0.005 Ry at $\omega=0$ and quadratically coarser at larger $\omega$
	   (Sec.II-D in Ref.\onlinecite{kotani_fusion_2010}).
	   The bin width becomes twiced at 0.04Ry. For integration along imaginary axis, we 
	   use ten points in the imaginary  axis of $\omega$. 
	   The parameters are good enough to give reasonable results as seen 
	   in Ref.\onlinecite{kotani_fusion_2010}.
\item[(E)]
	  $\EMAX$
\item[(F)]
	  Use ESs or not.
\item[(G)]
	  LDA or GGA, which are used as an assistance of numerical
	  calculation in PMT-QSGW.
	  See at the bottom of Sec.\ref{sec:ov}.
\end{enumerate}
Here (E),(F) and (G) are settings in common with the LDA/GGA-level calculations.

In Table \ref{tab:bandgapqsgw}, we show the
band gaps for GaAs and SiO2c calculated by PMT-QSGW for changing setting of (A)-(G).
We calculate the self-energy only at $\bfk$-mesh points, which are
$4\times4\times4$ and $2\times2\times2$ 
in the 1st BZ for GaAs and SiO2c, respectively
(we use large enough $\bfk$-mesh for electron density,
$10\times10\times10$ for GaAs, and $6\times6\times6$ for SiO2c).
No spin orbit coupling is included.
In the calculation of polarization function of \req{eq:polf},
we take all occupied and unoccupied states.
The top line date labeled as 'REF', which show the gaps 1.939 eV(GaAs) and 11.16 eV(SiO2c),
are treated as bases for following comparisons with other cases.
For the cases of 'REF', we take all bands ('all' for the column of $\EMAXS$ means taking
all the matrix elements of $\Dvxc$, that is, $\EMAXS$ is infinity).
Empty spaces in the Table \ref{tab:bandgapqsgw} mean that we use 
the same settings with the case of 'REF'.
For example, the next line to REF for GaAs means a case
with the same settings with REF except changes of
$|{\bf q+G}|^{\Psi}_{\rm Max}$=6.0 and $|{\bf q +G}|^{W}_{\rm Max}$=4.0.

We can see following points from the Table \ref{tab:bandgapqsgw}.
Generally speaking (as we see followings), it seems not so easy
to attain numerical error within $\sim$0.1 eV. 
Thus we take $\sim$0.1 eV as our target of numericall accuracy in the PMT-QSGW method.
It is not so meaningful to discuss about small differences.
\begin{enumerate}
\item
At the first section, we can see the dependence 
on $|{\bf q+G}|^{\Psi}_{\rm Max}$ and $|{\bf q +G}|^{W}_{\rm Max}$.
We see that the REF setting, $(|{\bf q+G}|^{\Psi}_{\rm Max}$,$|{\bf q +G}|^{W}_{\rm Max})
=$(4.0,3.0) (bohr)$^{-1}$, 
show convergence of  $\sim$0.01 eV even for the case of SiO2c
($\sim$ 0.001 eV for GaAs) for these parameters.
We have shown similar check in Ref.\onlinecite{kotani_quasiparticle_2007}.

\item
In our test cases of the PB in Table \ref{tab:pbset},
we can estimate numerical errors caused by the choice of PB.
As for PB in GaAs, 1.939 eV given by PB1 (REF) gives good agreement 
with 1.946 eV by PB1$l$, which is the largest PB among what we used here.
For SiO2c, we have little dependence on the choice of the PB used here. 
Especially, in case of PB0s, we use a set of PB on oxygen only with $l_{\rm cut}=2$.
This choice reduces the computational time so much for larger systems.

\item
The band gap gradually increases when we increase $\EMAX$ in GaAs.
The band gap monotonically changes from 1.939 eV at $\EMAX$=3.0 Ry to 1.969 eV at $\EMAX$=6.0 Ry for 'vwn'
(we see similar changes for 'vwn,es' where 1.945 eV to 1.982 eV).
Thus we can not see convergence behavior within this range of $\EMAX$.
This 1.969 eV can be taken as the best value for 'vwn' among performed
calculations in the sense of largest number of APWs.
Because of over-completeness problem of basis sets in the PMT method, 
it is not easy to enlarge number of APWs.
In addition, eigenfunctions at high energy are not accurate enough 
(we do not include local orbital for high energy bands).
Thus we inevitably takes this behevior as a limitation of our current
implementation of the PMT-QSGW.
Recall that such slow convergence on the number of unoccupied bands (= number
of APWs in our case) is also observed in Ref.\onlinecite{friedrich_band_2011}.

We observe similar behavior in the case of SiO2c. 
The band gap of SiO2c changes from 11.16 eV at $\EMAX$=3Ry (REF), 
to 11.18 eV at $\EMAX$=6.0 Ry. We see similar change for 'vwn,es';
it is from 10.49 eV at $\EMAX$=3.0 Ry to 10.41 eV at $\EMAX$=6.0 Ry. 

\item
Let us discuss other points for GaAs. 

At first, we see that using $\EMAXS$=3.0 Ry (marked by **) 
gives little difference from REF (1.942-1.939 eV).
Thus we may use $\EMAXS$=3.0 Ry to reduce computational efforts.

We should take ``vwn,es'' gives better values 
than ``vwn'' because we include the MTOs of ESs as bases.
We see the difference between 'vwn' and 'vwn,es' is small enough
(1.945-1.939=0.006 eV at $\EMAX$= 3.0 Ry; 
 1.982 -1.969 = 0.013 eV at $\EMAX$= 6.0 Ry).
Thus we do not need to use ESs for GaAs.

There are other cases where we have no clear explanations 
because kinds of factors can affects to results.
 In the case of ``vwn,es'', 1.945 eV (for $\EMAXS$='all') changes to
1.903 eV for $\EMAXS$= 3 Ry. Corresponding change in 'vwn'
is from 1.939eV to 1.942 eV. 

When we use 'pbe' as the assistance of numerical calculation
(explained at the bottom of \refsec{sec:ov}), result changes a little.
The best value 2.002 eV (marked by *) show a little 
difference of 0.02 eV from that in 'vwn,es' of 1.982 eV (marked by *).

As a conclusion, except non-converging behavior on $\EMAX$,
it might be safer to estimate numerical errors as $\sim$ 0.1 eV,
based on the dependence on computational conditions.


%

\item
Let us discuss other points for SiO2c.

In this case, we see not a small dependence on $\EMAXS$ for 'vwn';
it changes from 10.38 eV at $\EMAXS$=3.0Ry (marked by **) 
to 11.16 eV for 'all' (REF). The difference 11.16-10.38=0.78 eV looks too
large, much more than our target of numerical error $\sim$0.1 eV.
In 'vwn,es', corresponding values are 10.09 eV and 10.49 eV, respectively. 
We see that the difference 10.38-10.09 eV for $\EMAXS$=3.0Ry
between 'vwn' and 'vwn,es' is relatively small.
However, the difference becomes larger as 11.16-10.49 eV for
$\EMAXS$='all'. This means that the difference comes from the high
energy part of the matrix elements of $\Dvxc$. 
Generally speaking, higher energy parts are less reliable numerically.
Considering the fact of no MTOs in ES, 
we think 11.16 eV of REF is not so reliable.

As we see the above paragraph, 
it looks not easy to obtain convergence for $\EMAXS$ in this case.
Thus we think that we need to introduce a restriction to have good numerical accuracy.
For example, we may look for convergence for $\EMAXS$=3.0Ry. 
In fact, at $\EMAXS$=3.0Ry, the difference between 'vwn' and 'vwn,es' is
relativey small, 10.38-10.09 eV. That is, we can calculate
the QSGW band gap with the numerical error of
$\sim$ 0.3 eV for $\EMAXS$=3.0Ry . 
(in this case, 10.38 eV accidentally 
gives good agreement with 'vwn,es' for $\EMAX$=6.0Ry).


Note that the difference between 'vwn,es' and 'pbe,es'. It gives an extra
numerical error of $\sim$ 0.1eV.

\end{enumerate}

As a summary, convergence beheviors for band gap are satisfactory 
(convergence within $\sim$ 0.1 eV) except for $\EMAX$ when we include ESs.
This was not apparent in FP-LMTO-QSGW since we have 
no $\EMAX$ (no APWs). This is a limitation of 
the current implementation due to the limited ability of the PMT
method to describe high energy bands 
(overcompleteness problem of a basis set).
In addition, we see dependence on $\EMAXS$ when we do not use ESs 
in the case of SiO2c; including ESs is not convenient to treat
system such as slab models. If we use $\EMAXS$=3.0Ry, we have
smaller difference $\sim$ 0.3 eV from the case including ESs.

Considering the balance of computational efforts and accuracy,
we think that ``PMT-QSGW with $\EMAXS=\EMAX=3.0$ Ry without ESs''
or similar is useful for practical calculations.
This is taken as an approximation to 
the exact results of the fundamental equation of QSGW.

It might be not so meaning to obtain fully converged results in QSGW,
since it is inevitable for QSGW to give some differences from
experimental values. In fact, QSGW tends to give a little too large band gaps
\cite{vans06,kotani_quasiparticle_2007} even if it is accurately
performed. For example, calculated value of 10.41 eV (vwn) for SiO2c in 
Table \ref{tab:bandgapqsgw} is rather larger than the experimental value $\sim$ 8.9 eV
\cite{sio2cgap}, thus not directly compared with experiments
(Other QSGW calculation by Shaltat et al \cite{shaltaf_band_2008} 
gives band gap 8.8 eV by QSGW for SiO2c. The difference from our value
of 10.41 eV may indicate numerical difficulty to have convergence).
In cases, we need to correct this discrepancy from experiments 
empirically by a hybrid method such as
$(1-\alpha)\times$QSGW+$\alpha\times$LDA as was used 
in Refs.\onlinecite{chantis_ab_2006,kotani_impact_2010}
when we like to have good agreement with experiments ($\alpha \sim$ 0.2).
Thus, from a practical point of view, it will be better to
take the parameter $\alpha$ as a combined correction on the
theoretical error and the numerical errors due to 
the approximation as `PMT-QSGW with $\EMAXS=\EMAX=3.0$ Ry without ESs''.
Or we may need to invent a better fundamental equation to go beyond QSGW,
which is numerially stable with keeping advantages of QSGW and give
better correspondence with experiments.


\begin{figure}[ht]
\caption{\label{fig:nkconv}
Dependence of the band gap as for the number of $\bfk$ 
points in the 1st BZ for self-energy calculation.
Integer $n$ of x-axis means the number of division of BZ is $n \times n \times n$.
y-axis means the band gap. }
\begin{flushleft}
\includegraphics[width=8.5cm]{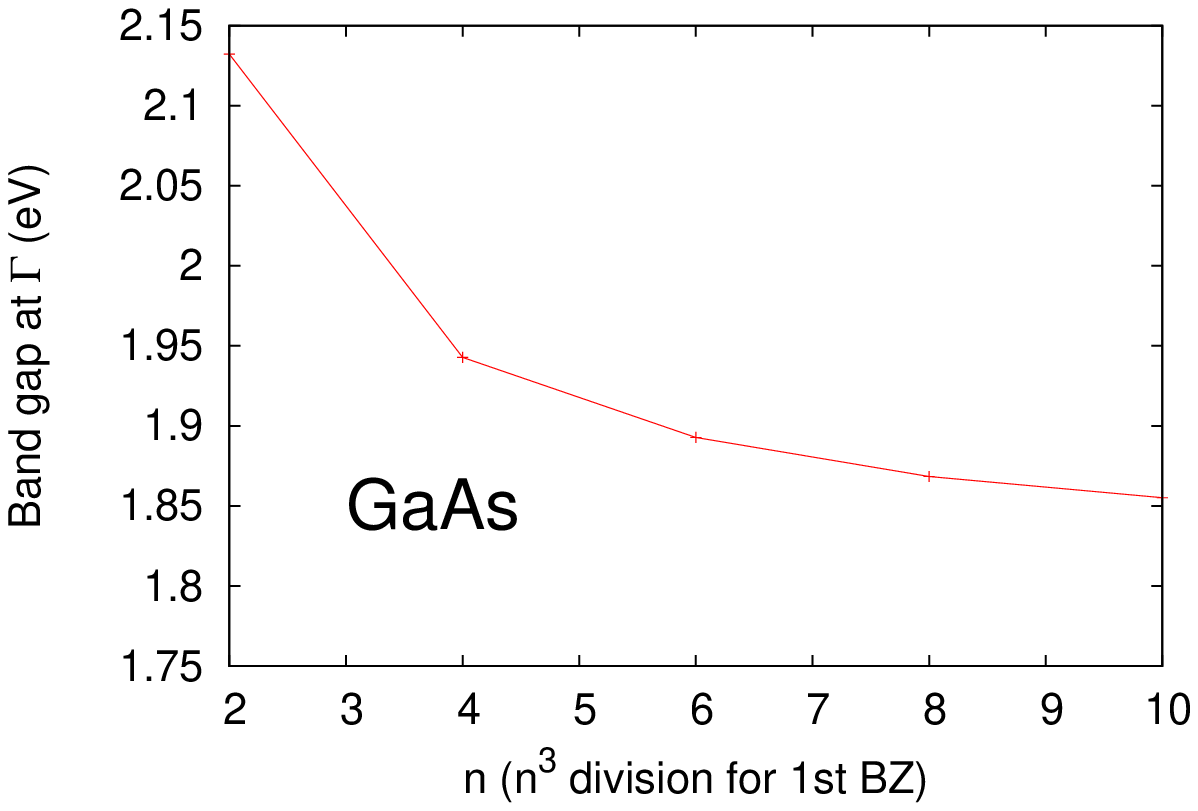}
\includegraphics[width=8.5cm]{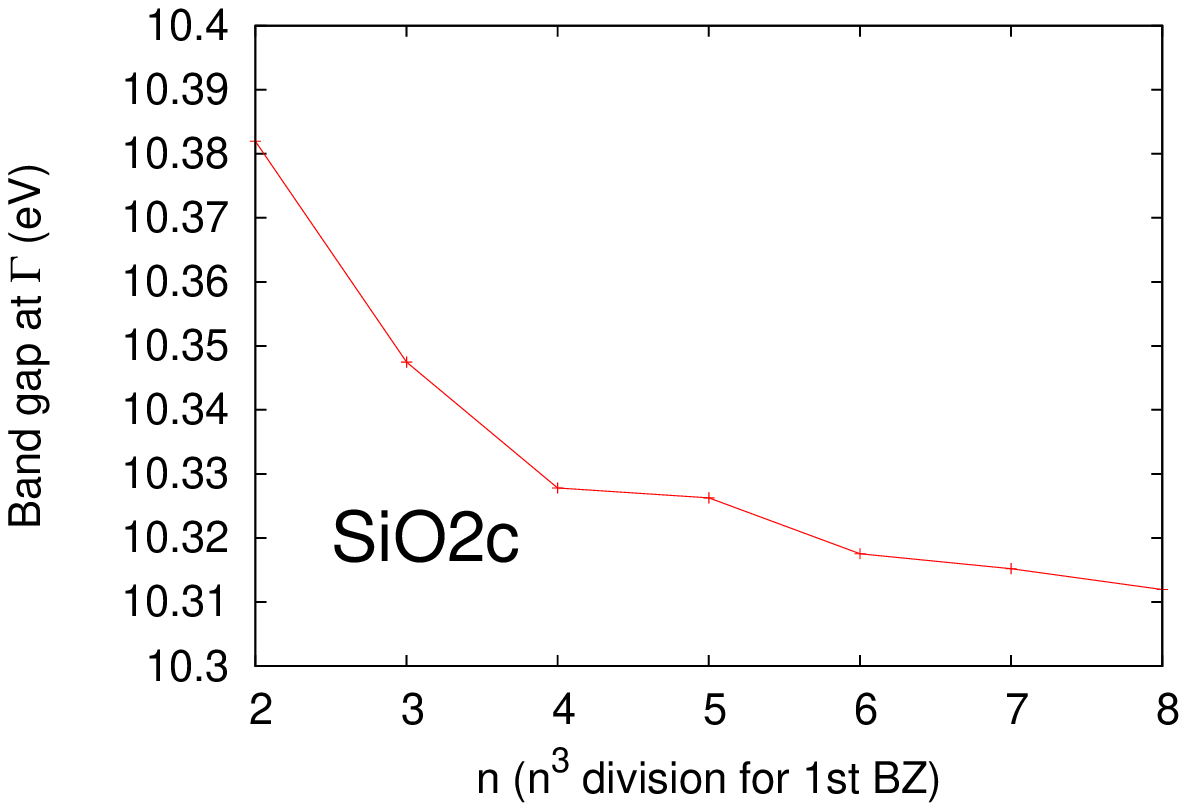}
\end{flushleft}
\end{figure}
\begin{figure}[ht]
\caption{\label{fig:qsgwband}
Band plot for GaAs, corresponding to the case of $10\times10\times10$ in Fig.\ref{fig:nkconv},
and for SiO2c to the case of $8\times8\times8$.}
\begin{flushleft}
\includegraphics[width=8.5cm]{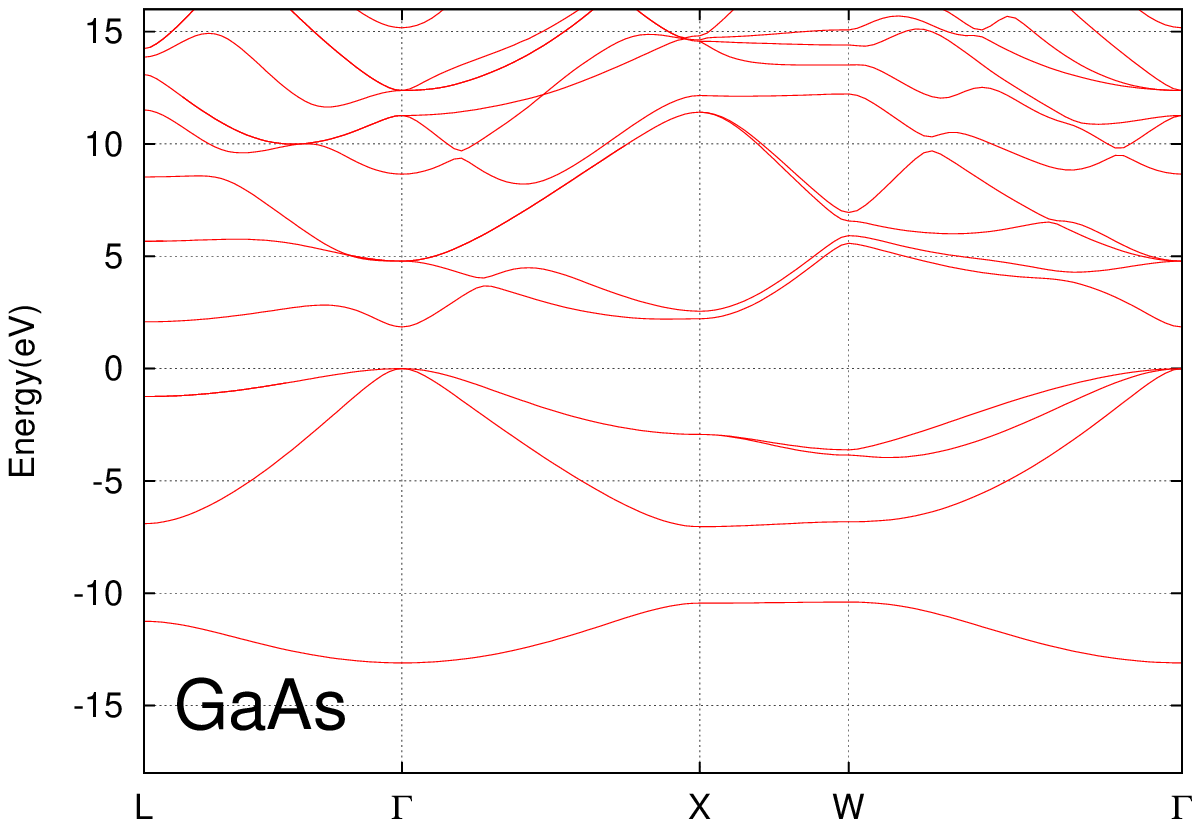}
\includegraphics[width=8.5cm]{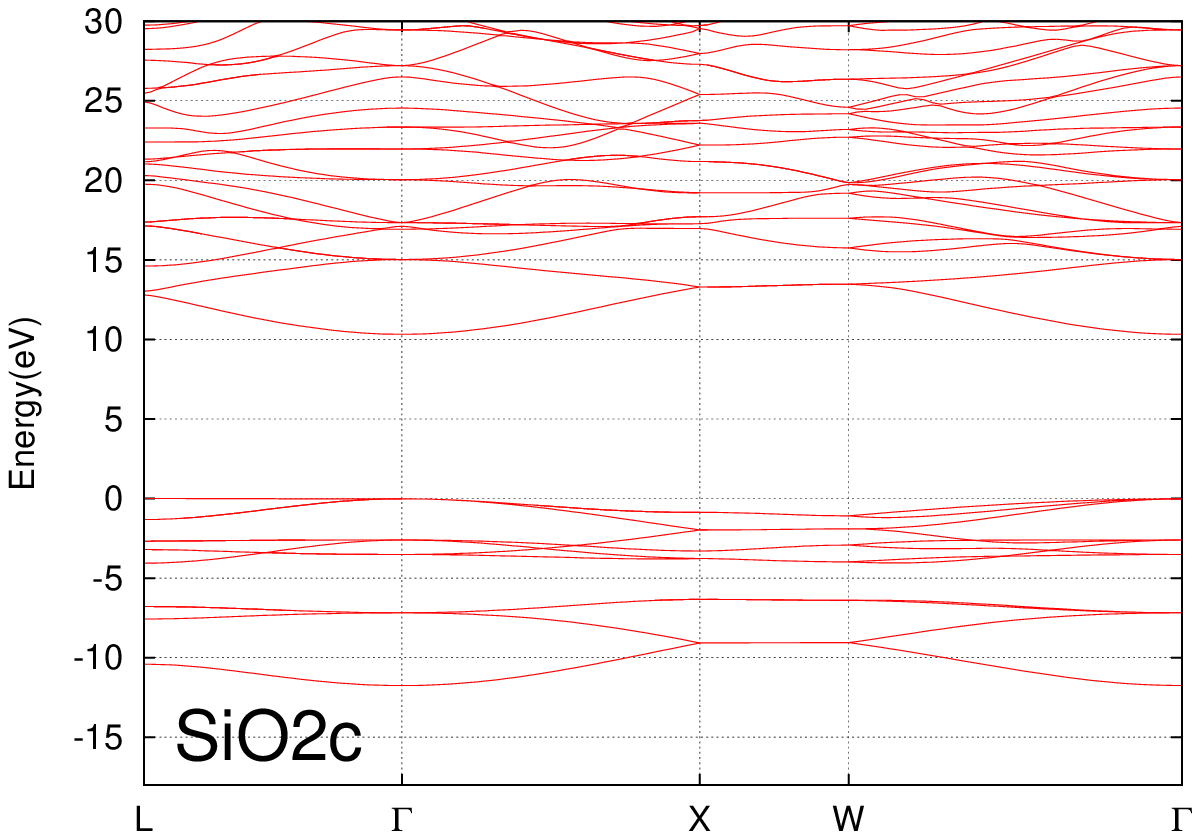}
\end{flushleft}
\end{figure}
In Fig.\ref{fig:nkconv}, we show the convergence check 
about the number of $\bfk$ points for self-energy calculation in the 1st
BZ (the $\bfk$ point mesh for electron density is fixed).
The integer $n$ of x-axis means that 
the used number of $\bfk$ points is $n\times n\times n$.
As for GaAs, we see smooth convergence on the number of $\bfk$ points.
In the $4\times 4 \times 4$ calculation,
we see $\sim$ 0.1eV overestimation in comparison with the value 
at $10 \times 10 \times 10$. We need to choose number of $\bfk$ points,
to have best accuracy within the allowed computational resources.
As for SiO2c, pay attention to the energy scale of y-axis.
The difference of the gap between $n=2$ and $n=8$ is rather small, 
only $\sim$0.04 eV. 
In our analysis, unsmooth behavior of this plot
is because of the cutoff of $\EMAXS$; see the dependence on $\EMAXS$ 
in Table.\ref{tab:bandgapqsgw}. Energy bands near $\EMAXS$ are
taken into account or not by a slight change of $\bfk$ point.

In Fig.\ref{fig:qsgwband}, the energy dispersion curve for \QSGW\
obtained with the largest number of $\bfk$ point cases in Fig.\ref{fig:nkconv}
are shown, in order to show the difference from LDA/GGA.

\section{summary}
We have developed a new method, the PMT-QSGW method
to perform the QSGW calculation based on the PMT method. 
PMT-QSGW have advantages in the robustness, easy to use, and accuracy in
comparison with FP-LMTO-QSGW. We do not need to tune parameters for MTOs.
Thanks to APWs, we can use highly localized MTOs
with low energy APWs ($\sim$ 4 Ry). 
Then we employ simplified interpolation procedure to the static component of the self-energy
instead of previous complicated one in FP-LMTO-QSGW.

We have shown detailed convergence check on the band gaps of two typical cases,
GaAs and cubic SiO$_2$. We analyzed how their band gaps depend
on the cutoff parameters and computational settings. 
Then we see the performance and limitations of PMT-QSGW.
We suggest ``PMT-QSGW with $\EMAXS=\EMAX=3.0$ Ry without ESs''
as an approximaion for practical usage.
Results shown in this paper can be reproduced by the PMT-QSGW method implemented in the 
\texttt{ecalj} package, which is freely available from github \cite{ecalj}.

The PMT-QSGW method with the highly localized MTOs and low energy APWs
is advantageous for theoretical treatment. 
Techniques developed here can be useful even to go beyond QSGW.

\  \\

Acknowledgement:\\

I thank to Dr.H.Kino for discussions, codings, and advises to this manuscript. 
This work was partly supported by Advanced Low Carbon Technology
Research and Development Program (ALCA) of Japan Science and Technology
Agency (JST), and by Grant-in-Aid for Scientific Research 23104510. 
We also acknowledge computing time provided by Computing System for
Research in Kyushu University.
\bibliography{refsk}
\end{document}